\documentclass[preprint,authoryear,12pt]{elsarticle}
\usepackage{lscape, morefloats, graphicx, float}   
\usepackage{graphicx}
\RequirePackage{color}                   

\begin{document}

\begin{frontmatter}

\title{Vertical and radial metallicity gradients in high latitude galactic fields with SDSS}

\author[1]{Sabiha Tun{\c c}el G\"u{\c c}tekin\corref{cor}}\ead{sabihatuncel@gmail.com}
\author[2]{Sel\c cuk Bilir}\ead{sbilir@istanbul.edu.tr}
\author[2]{Salih Karaali}\ead{karsa@istanbul.edu.tr}
\author[1]{Olcay Plevne}\ead{olcayplevne@gmail.com}
\author[2]{Serap Ak}\ead{akserap@istanbul.edu.tr}

\address[1]{Istanbul University, Graduate School of Science and Engineering, 
Department of Astronomy and Space Sciences, 34116, Beyaz\i t, Istanbul, Turkey}
\address[2]{Istanbul University, Faculty of Science, Department 
of Astronomy and Space Sciences, 34119 University, Istanbul, Turkey}

\begin{abstract} 
We used the $ugr$ magnitudes of 1,437,467 F-G type main-sequence stars with metal abundance $-2\leq[{\rm Fe/H}]\leq+0.2$ dex and estimated radial and vertical metallicity gradients for high Galactic-latitude fields, $50^{\circ}< b\leq 90^{\circ}$ and $0^{\circ}< l\leq 360^{\circ}$, of the Milky Way Galaxy. The radial metallicity gradient $d[{\rm Fe/H}]/dR=-0.042\pm0.011$ dex kpc$^{-1}$ estimated for the stars with $1.31< z\leq1.74$ kpc is attributed to the thin-disc population. While, the radial gradients evaluated for stars at higher vertical distances are close to zero indicating that the thick disc and halo have not undergone a radial collapse phase at least at high Galactic latitudes. The vertical metallicity gradients estimated for stars with three different Galactic latitudes, $50^{\circ}< b\leq 65^{\circ}$, $65^{\circ}< b\leq 80^{\circ}$ and $80^{\circ}< b\leq 90^{\circ}$ do not show a strong indication for Galactic latitude dependence of our gradients. The thin disc, $0.5<z\leq 2$ kpc, with a vertical metallicity gradient $d[{\rm Fe/H}]/dz= -0.308\pm0.018$ dex kpc$^{-1}$, is dominant only in galactocentric distance interval $6<R\leq 10$ kpc, while the thick disc ($2<z\leq 5$ kpc) could be observed in the intervals $6<R\leq 10$ and $10<R\leq 15$ kpc with compatible vertical metallicity gradients, i.e. $d[{\rm Fe/H}]/dz= -0.164\pm0.014$ dex kpc$^{-1}$ and $d[{\rm Fe/H}]/dz= -0.172\pm0.016$ dex kpc$^{-1}$. Five vertical metallicity gradients are estimated for the halo ($z>5$ kpc) in three galactocentric distance intervals, $6<R\leq 10$, $10<R\leq 15$ and $15<R\leq 20$ kpc. The first one corresponding to the interval $6<R\leq 10$ kpc is equal to $d[{\rm Fe/H}]/dz= -0.023\pm0.006$ dex kpc$^{-1}$, while the others at larger galactocentric distances are close to zero. We derived synthetic vertical metallicity gradients for 2,230,167 stars and compared them with the observed ones. There is a good agreement between the two sets of vertical metallicity gradients for the thin disc, while they are different for the thick disc. For  the halo, the conspicuous difference corresponds to the galactocentric distance interval $6<R\leq 10$ kpc, while they are compatible at higher galactocentric distance intervals. 
\end{abstract}


\end{frontmatter}
\parindent=0.5 cm

\section{Introduction}
Metallicity is an important tool used for understanding the formation and evolution of our Galaxy. \citet*{ELS62} proposed the first model for the formation and evolution of the Milky Way. In this model, the first stars are formed from the metal-poor protogalactic cloud. The heavier elements produced by  massive stars are spread to the interstellar medium with supernova explosions and the second generation of stars formed from relatively metal-rich molecular clouds, which will proceed to collapse of the matter to the Galactic plane and eventually become the disc component. According to the model, this collapse process happened in a short time scale (a few of $10^8$ yr) and this might cause a metallicity gradient in the Galaxy \citep*[cf.][]{Andrievsky02, Karaali04, Bilir06}. Over time, understanding that the halo global clusters have different ages and different chemical abundances has led \citet{Searle78} to suggest an alternative model. In this model, the lack of a metallicity gradient in the halo indicates that the Milky Way was formed by the accretion of clouds with different chemical composition over a long time scale (several of $10^9$ yr). 

Vertical metallicity gradients in the Milky Way are deeper than the radial ones, and their numerical values depend on the observed object itself as well as on its position. The numerical values in \citet{Ak07a}, $d[{\rm Fe/H}]/dz=-0.38\pm0.06$ dex kpc$^{-1}$, and in \citet{Huang15}, $d[{\rm Fe/H}]/dz=-0.052\pm0.010$ dex kpc$^{-1}$ can be given as two extreme values for the vertical metallicity gradients. The first one was estimated for G spectral type stars in the distance interval $3\leq z < 5$ kpc while the second one belongs to the red clump (RC) stars observed in the distance intervals $|z|\leq1.9$ kpc and $9<R\leq10$ kpc. The deepest radial metallicity gradient is the one in \citet{Frinchaboy13}, i.e. $d[{\rm Fe/H}]/dR=-0.20\pm0.08$ dex kpc$^{-1}$, estimated for open clusters with galactocentric distance $7.9<R<10$ kpc. The radial metallicity gradients in \citet{Xiang15}, $d[{\rm Fe/H}]/dR=-0.100\pm0.003$ dex kpc$^{-1}$, and in \citet{Onal16}, $d[{\rm Fe/H}]/dR=-0.047\pm0.003$ dex kpc$^{-1}$ represent the intermediate and relatively low radial metallicity gradients in the literature. The first gradient was estimated for the turnoff stars with distance $|z|\leq0.1$ kpc and age $2<t<16$ Gyr while the second one belongs to RC stars with $|z|\leq0.5$ kpc. A large reference table related to the metallicity gradients can be found in \citet{Onal16}. Positive metallicity gradients which are usually estimated for stars at relatively large vertical and radial distances \citep[cf.][]{Cheng12} confirm the argument that some of the components of our Galaxy are formed from merger and aggregation of numerous fragments \citep{Searle78, Freeman02}.

Metallicity of a star can be determined spectroscopically or photometrically. The first procedure requires high resolution which is available for nearby dwarf and giant stars at different distances. While the second one can be applied to distant stars as well. The distance of a star relative to the Sun is necessary for metallicity gradient estimation which can be determined by using trigonometric or photometric parallaxes. Trigonometric parallax can be estimated only for nearby stars and one of the main sources is the {\em Hipparcos} catalogue \citep{ESA97, vanLeeuwen07}. Another source is the {\em Tycho}-{\em Gaia} Astrometric Solution (TGAS) catalogue of \citet*{Michalik15} which covers the trigonometric parallaxes (and proper motions) of about $2.5\times10^6$ stars supplied from the {\em Gaia} Data Release 1 \citep{Lindegren16}. Photometric parallax of a star can be provided by a combination of its apparent and absolute magnitude \citep*[cf.][]{Bilir05, Bilir08a, Bilir09}, where absolute magnitude needs also to be determined. The procedure usually used for this purpose is based on its offset from the fiducial sequence of a cluster such as Hyades \citep{Laird88, Nissen91, Karaali03a, Karaali03b, Karaali05, Karaali11, Karatas06}. Alternative methods can be found in \citet{Phleps00}, \citet{Chen01}, \citet{Siegel02} and \citet{Ivezic08}.      

In \citet[][hereafter Paper I]{Guctekin16}, we calibrated the iron abundance [Fe/H] and absolute magnitude offset $\Delta M_V$ in terms of the  ultraviolet (UV) excess $\delta_{0.6}(U-B)$, while in \citet[][hereafter Paper II]{Guctekin17} we carried out similar calibrations for the $ugr$ photometric data of the Sloan Digital Sky Survey \citep[SDSS,][]{York00}. In this study we aim to test the mentioned calibrations in Paper II on stars in a large, 5,280 deg$^2$, high Galactic latitude, $b>50^{\rm o}$, star field. Thus, we should obtain vertical and radial metallicity gradients, $d[{\rm Fe/H}]/dz$ and $d[{\rm Fe/H}]/dR$ respectively, in this direction of our Galaxy and compare them with the counterparts. The paper is organized as follows: the data are presented in Section 2, estimations of the metallicity gradients are given in Section 3, and finally Section 4 is devoted to the summary and discussion.  

\section{Data}
\subsection{The selection of the F-G type main-sequence stars}
The de-reddened $ugr$ magnitudes used in this study are provided from the recent survey DR12 of SDSS III \citep{Alam15}. There are 26,719,324 objects  with de-reddened $ugriz$ magnitudes in an area of 5,280 deg$^2$ centred on the north Galactic pole. The $ugr$ magnitudes and their errors are taken from the SQL webpage of SDSS\footnote{https://skyserver.sdss.org/dr12/en/tools/search/sql.aspx}. We applied a series of constraints, as explained in the following, and obtained a sample of F-G spectral type dwarfs suitable to estimate vertical and radial metallicity gradients. We applied the procedure of \citet{Chen01}, i.e. $g_0\leq23$ and $(u-g)_0>0.5$ mag, by doing so the extragalactic objects such as quasars are omitted. Thus, the original set of objects is reduced to 16,281,265 stars. We noticed that there are large scattered objects in the $(g-r)_0 \times (r-i)_0$ two-colour diagram (Fig. 1). Hence, we adopted the following equation of \citet{Juric08} and limited the number of stars to 15,456,605 which lie within $\pm 2\sigma$ of this equation:
\begin{eqnarray}
(g-r)_0 = 1.39 (1-\exp[-4.9(r-i)_0^3-2.45(r-i)_0^2 \\ \nonumber
 -1.68(r-i)_0-0.050]).  
\end{eqnarray} 

\begin{figure}[t]
\begin{center}
\includegraphics[scale=0.33, angle=0]{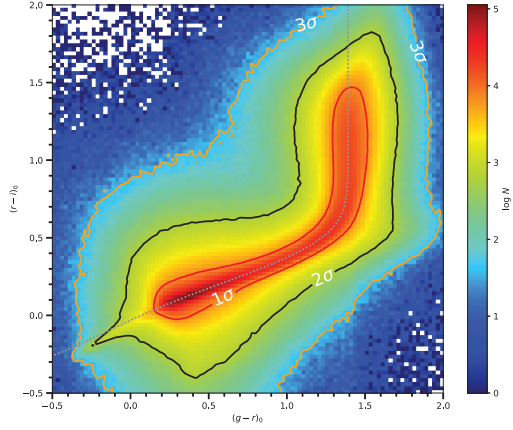}
\caption[] {Two-colour diagram for 16,281,265 stars. The dotted mean line corresponds to the equation of \citet{Juric08}, while the lines with red, black and orange colours cover the stars within $1\sigma$, $2\sigma$ and $3\sigma$ of the mean line, respectively.}
\end{center}
\end{figure}

\begin{figure*}[t]
\begin{center}
\includegraphics[scale=0.38, angle=0]{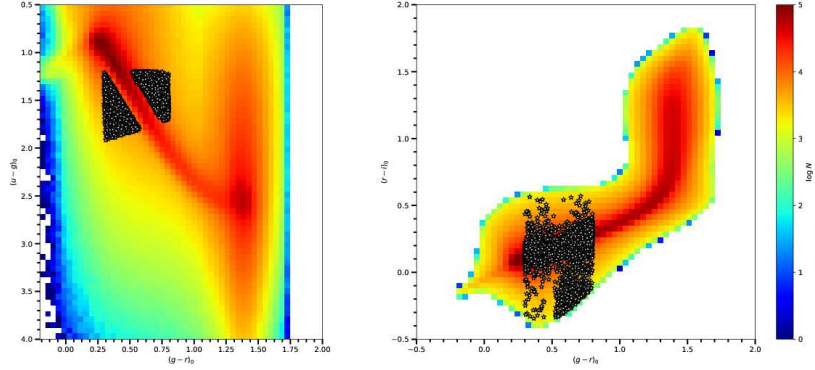}
\caption[] {Two-colour diagrams for 15,448,992 stars within $2\sigma$ of the mean line. 7,613 gaint stars are represented with black asterisk symbol in both diagrams.}
\end{center}
\end{figure*}

The next step concerns the giants in this field which are identified by the following set of equations of \citet[][ see also \citet{Bilir08b}]{Helmi03}:

\begin{eqnarray}
1.1\leq(u-g)_{0}\leq2, \nonumber\\
-0.1<P_1<0.6,\\ \nonumber
|s|>m_{s}+0.05.
\end{eqnarray}
where $P_1$, $s$ and $m_s$ are defined as follows:
\begin{eqnarray}
P_{1}=0.910(u-g)_{0}+0.415(g-r)_{0}-1.28,\nonumber\\
s=-0.249u_{0}+0.794g_{0} - 0.555r_{0}+0.240, \\ \nonumber
m_s=0.12. 
\end{eqnarray} 
         
This constraint revealed 7,613 giants which reduced our set of stars to 15,448,992 dwarfs (Fig. 2). The colour range of the F-G spectral type main-sequence stars, $0.15<(g-r)_0<0.50$ mag, further decreased the number of dwarfs down to 4,402,895. The additional constraint on the apparent magnitude, $g_0\leq 20$, to exclude stars with relatively large errors reduced the number of dwarfs to 1,895,264. The brighter limit for the apparent magnitude is $g_0=14$ mag. Finally, we omitted the dwarfs whose metallicities lie out of the metallicity range of our calibration obtained in Paper II, i.e. $-2\leq[{\rm Fe/H}]\leq 0.2$ dex. Thus, the final F-G spectral type dwarf sample consists of 1,437,467 stars. The magnitude and colour errors for these stars are as follows: $g_{err}= 0.006$, $(u-g)_{err}=0.022$, $(g-r)_{err}=0.008$, and $(r-i)_{err}=0.008$ mag.

\subsection{Metallicities, absolute magnitudes and distances}
The [Fe/H] metallicities and $M_g$ absolute magnitudes for the (final) sample stars are estimated by using the calibrations obtained in Paper II. For the metallicity we used the following equation:
\begin{eqnarray}         
[{\rm Fe/H}]=0.105(0.010)-5.633(0.521)\times \delta_{0.41}\\ \nonumber
+2.984(1.895)\times \delta_{0.41}^2-27.209(16.359)\times \delta_{0.41}^3.\\ \nonumber
\end{eqnarray}
where $\delta$ is the ultra-violet excess of a star relative to a Hyades star with the same colour reduced to $(g-r)_0=0.41$ mag. While for the absolute magnitude, we estimated an offset $\Delta M_g$ for a given star by the following equation and added it to the absolute magnitude of a Hyades star with the same $(g-r)_0$ colour: 

\begin{eqnarray}         
\Delta M_g=-0.133(0.031)+5.169(0.965)\times \delta_{0.41}\\ \nonumber
-3.623(0.978)\times \delta_{0.41}^2+21.497(7.799)\times \delta_{0.41}^3.\\ \nonumber        
(M_g)_{*}= \Delta M_g+(M_g)_H.\\ \nonumber
\end{eqnarray}
where $(M_g)_*$ and $(M_g)_H$ correspond to the absolute magnitudes of the star in question and that of the Hyades star with the same $(g-r)_0$ colour. 

The distance $d$ of a star relative to the Sun is estimated by means of the Pogson formula, $g_0-M_g=5\log d-5$, while its distance to the Galactic plane, $z$, is evaluated via the equation $z=d\times \sin b$ where $b$ is the Galactic latitude of the star in question. Finally, radial distances ($R$) of the sample stars are provided by the following equation; 

\begin{eqnarray}        
R=\sqrt{R_0^2+d^2\cos^2 b-2R_0d\cos b\cos l+{{d^2\sin^2 b}\over{\eta^2}}}.
\end{eqnarray}
where $l$ and $b$ are the Galactic longitude and latitude of the related star, respectively, and where $R_0=8$ kpc is the distance to the Galactic centre \citep{Majewski93}, and $\eta$ is the oblate parameter of the halo whose numerical value is usually adopted as $\eta=0.85$ \citep{Carney96}. We used the same eqaution for the disc stars to avoid any discontinuty in the metallicity distribution versus distances (see Section 3). However we should note that the differences of the radial distances estimated for a spherical and oblate disc galaxy is less than 0.06 kpc.  
                        
\section{Metallicity gradients}
\subsection{Radial metallicity gradients}
We estimated radial metallicity gradients for 14 sub-samples defined by distances relative to the Galactic mid-plane, as explained in the following. First, we separated the sample stars into 18 Galactic longitude intervals with equal scale, i.e. $0^{\rm o} < l \leq 20^{\rm o}$, $20^{\rm o} < l \leq 40^{\rm o}$, ..., $320^{\rm o} < l \leq 340^{\rm o}$, $340^{\rm o} < l \leq 360^{\rm o}$ and re-separated each of these intervals into 14 sub-samples with distances relative to the Sun, $0.5 < d\leq 1$, $1<d\leq1.5$, $1.5<d\leq 2$, $2<d\leq 3$, $3<d\leq 4$, $4<d\leq 5$, $5<d\leq 6$, $6<d\leq 7$, $7<d\leq 8$, $8<d\leq 9$, $9<d\leq 10$, $10<d\leq 12$, $12<d\leq 14$ and $14<d\leq 16$ kpc. We performed metallicity histograms for each of these sub-samples and determined their modes. Also, we estimated the median values of the radial ($\widetilde{R}$) and vertical ($\widetilde{z}$) distances for the stars in these sub-samples. We present the metallicity histogram for the Galactic longitude interval $160^{\rm o} < l \leq 180^{\rm o}$ in Fig. 3 and the corresponding derived data just mentioned in Table 1 as an example. Second, we plotted the metallicity modes determined for each of the 14 sub-samples against radial distances ($\widetilde{R}$) and estimated the radial metallicity gradient for each of the mentioned sub-samples. The results are given in Fig. 4. The radial metallicity gradients in three vertical distance intervals corresponding to short distances, i.e. $0.44<z\leq0.87$, $0.87<z\leq 1.31$ and $1.31<z\leq1.74$ kpc, are $d[{\rm Fe/H}]/dR=-0.010\pm0.022$, $d[{\rm Fe/H}]/dR=-0.028\pm0.013$, and $d[{\rm Fe/H}]/dR= -0.042\pm0.010$ dex kpc$^{-1}$, while for distances larger than $z=1.74$ kpc three of the radial metallicity gradients are rather small negative numbers and eight of them are positive (Fig. 5). The $z$-intervals just mentioned  correspond to the intervals $0.5<d\leq 1$, $1<d\leq 1.5$, $1.5<d\leq 2$ kpc, and the $z$ values in Fig. 5 are taken from the fifth column of Table 1 which correspond to the $d$-values in the first column of the same table. The spatial distribution of the sample stars are given in the $z-R$ plane in three panels in Fig. 6 with colour-coded for number of stars ($N$), the distance relative to the Sun ($d$) and the metallicity ([Fe/H]).    

\begin{figure}[p]
\begin{center}
\includegraphics[scale=0.23, angle=0]{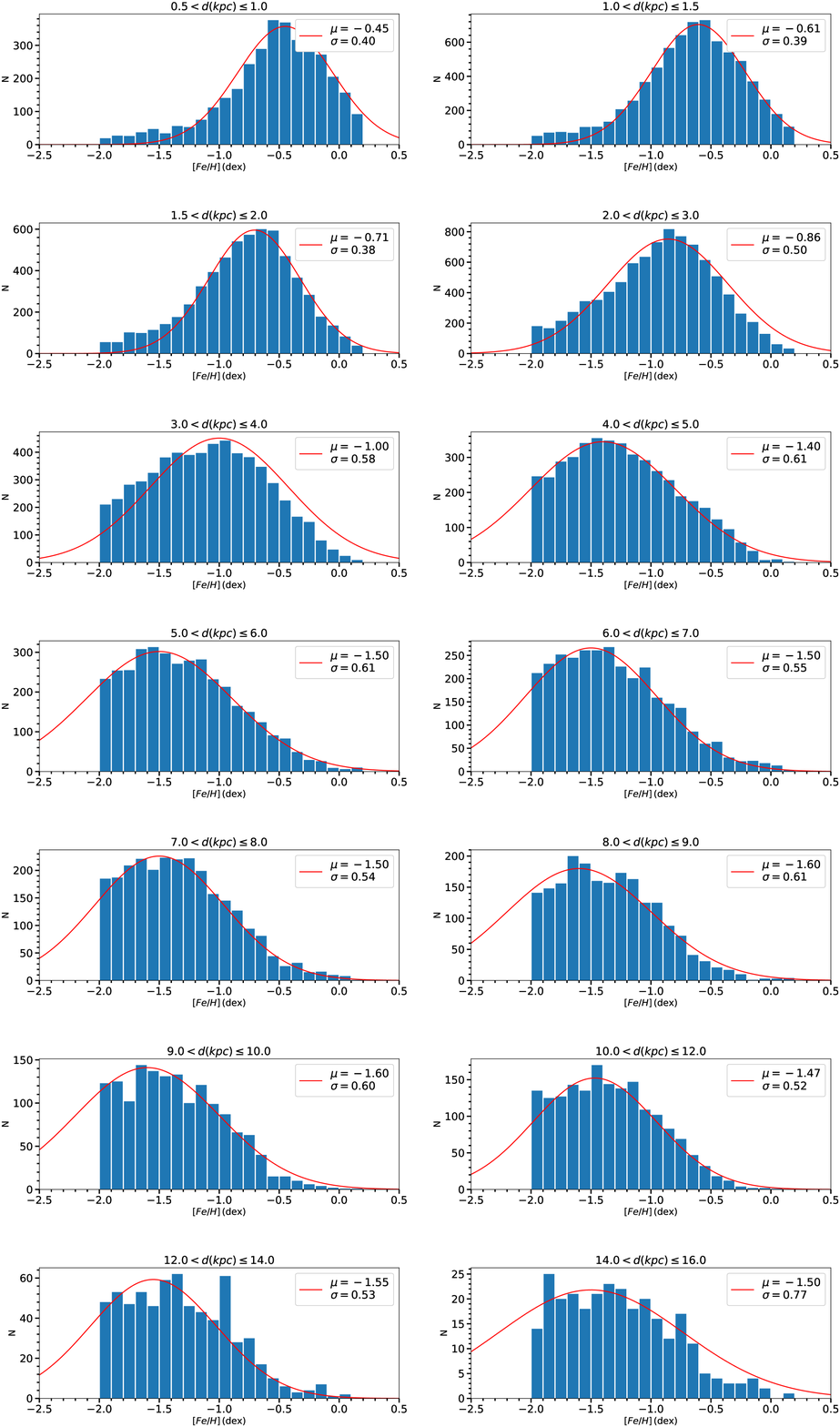}
\caption[] {Metallicity histograms for stars with Galactic longitudes $160^{\rm o}<l\leq 180^{\rm o}$. The thin line denotes the Gaussian curve whose mode is fitted to the one of the metallicity distribution. The Gaussian curve deviates gradually from the metallicity distribution at distances larger than $d=4$ kpc.} 
\end{center}
\end{figure}

\begin{figure}[p]
\begin{center}
\includegraphics[scale=0.19, angle=0]{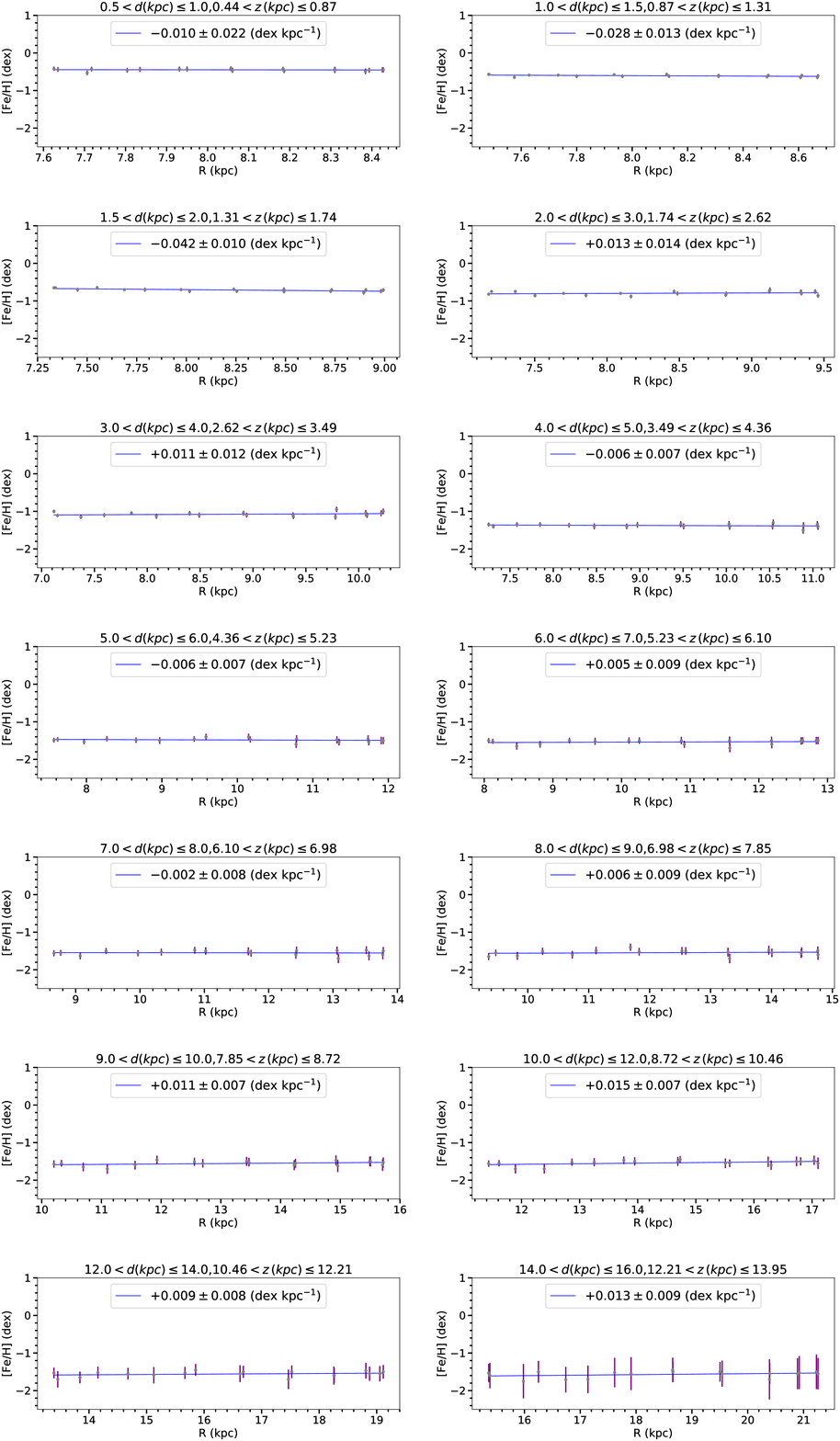}
\caption[] {Metallicity distribution with respect to the radial distance for stars in 14 sub-samples defined in the text and given in Table 1. Each panel covers the metallicities of stars in 18 Galactic longitude intervals defined in the text. The corresponding  metallicity gradients are indicated in each panel. The blue line corresponds to the metallicity calibration in terms of radial distance, while the red bar denotes the metallicity uncertainties}
\end{center}
\end{figure}

\begin{table}[!htb]
\setlength{\tabcolsep}{3pt}
\centering
\caption{Data for the stars with Galactic longitudes $160^{\rm o}<l\leq 180^{\rm o}$. The columns give: ($d_1-d_2$) distances relative to the Sun, ($z_1-z_2$) vertical distances, ($N$) number of stars, ($\widetilde{R}$) and ($\widetilde
{z}$) median radial and vertical distances, respectively, and ([Fe/H]) metallicity modes with their uncertainties.}
\begin{tabular}{cccccc}
\hline
$d_1-d_2$ & $z_1-z_2$ & $N$ &$\widetilde{R}$&$\widetilde{z}$& [Fe/H]\\
(kpc) & (kpc) &   & (kpc) & (kpc) & (dex) \\
\hline
0.5-1 & 0.44-0.87   & 3420 &  8.43 &  0.74 & $-$0.45$\pm$0.01 \\
1-1.5 & 0.87-1.31   & 7094 &  8.67 &  1.09 & $-$0.61$\pm$0.00 \\
1.5-2 & 1.31-1.74   & 6019 &  8.99 &  1.51 & $-$0.71$\pm$0.00 \\
2-3   & 1.74-2.62   & 8866 &  9.46 &  2.15 & $-$0.86$\pm$0.01 \\
3-4   & 2.62-3.49   & 5890 & 10.23 &  3.03 & $-$1.00$\pm$0.01 \\
4-5   & 3.49-4.36   & 4404 & 11.06 &  3.92 & $-$1.40$\pm$0.01 \\
5-6   & 4.36-5.23   & 3679 & 11.91 &  4.85 & $-$1.50$\pm$0.01 \\
6-7   & 5.23-6.10   & 3143 & 12.86 &  5.72 & $-$1.50$\pm$0.01 \\
7-8   & 6.10-6.98   & 2623 & 13.78 &  6.64 & $-$1.50$\pm$0.01 \\
8-9   & 6.98-7.85   & 2027 & 14.76 &  7.49 & $-$1.60$\pm$0.01 \\
9-10  & 7.85-8.72   & 1525 & 15.71 &  8.47 & $-$1.60$\pm$0.02 \\
10-12 & 8.72-10.46  & 1745 & 17.04 &  9.67 & $-$1.47$\pm$0.01 \\
12-14 & 10.46-12.21 &  665 & 19.06 & 11.60 & $-$1.55$\pm$0.02 \\
14-16 & 12.21-13.95 &  280 & 21.23 & 13.25 & $-$1.50$\pm$0.05 \\
\hline
\end{tabular}
\end{table}

\begin{figure}[!b]
\begin{center}
\includegraphics[scale=1.3, angle=0]{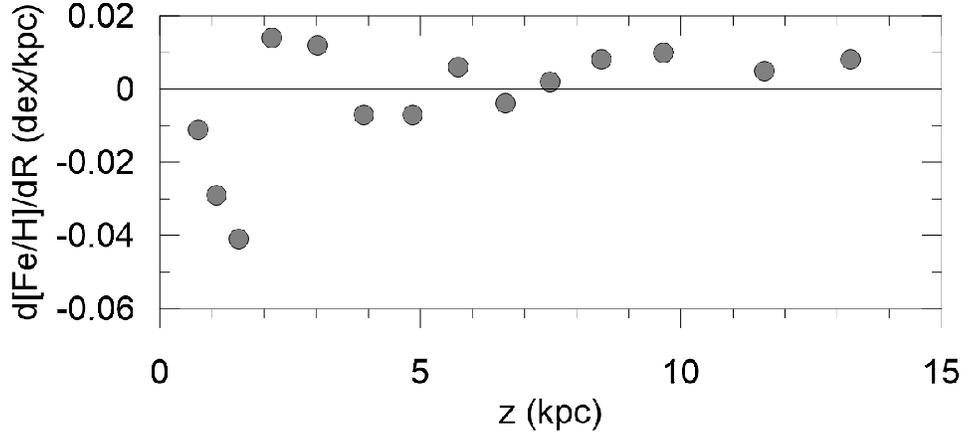}
\caption[] {Distribution of the radial metallicity gradients relative to vertical distance, for stars in the 14 sub-samples.}
\end{center}
\end{figure}

\begin{figure}[!b]
\begin{center}
\includegraphics[scale=0.43, angle=0]{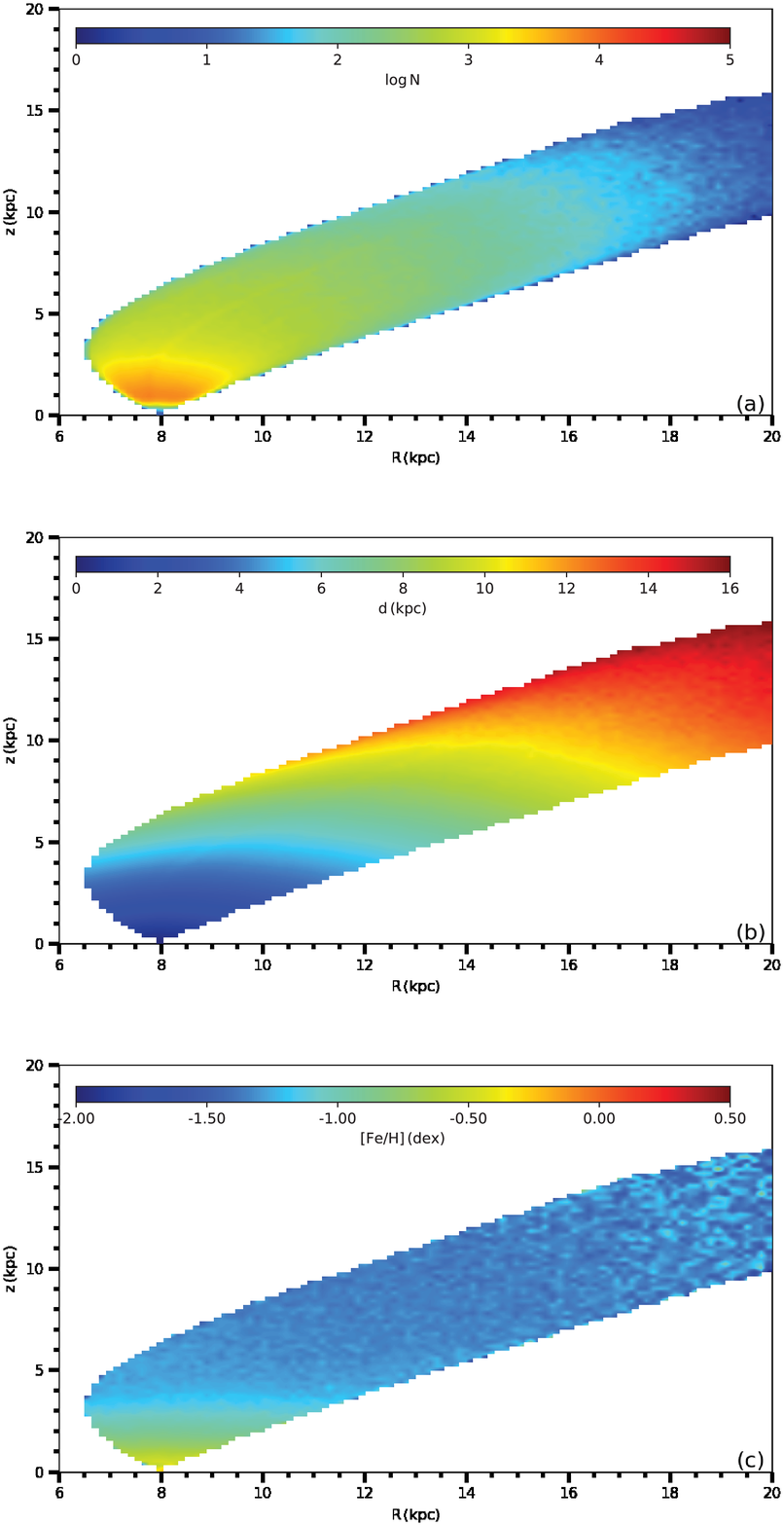}
\caption[] {Distribution of the sample stars in the $z-R$ plane in three panels, colour-coded for the number of stars (a), the distance relative to the Sun (b), and the metallicity (c).}
\end{center}
\end{figure}

\subsection{Vertical metallicities}
The vertical metallicity gradients are estimated for thin disc, thick disc, and halo components of the Galaxy separately. We adopted the galactocentric distance ranges $6<R<10$, $10<R<15$, and $15<R<20$ kpc and carried out this analysis for each range for stars with $0.5<z\leq2$ (thin disc), $2<z\leq5$ (thick disc), and $z>5$ (halo) kpc, by using two procedures  explained in the following. In the first procedure, the stars are separated into a series of sub-samples by using the three Galactic latitude and four Galactic longitude intervals: $50^{\circ}<b\leq65^{\circ}$, $65^{\circ}<b\leq80^{\circ}$, $80^{\circ}<b\leq90^{\circ}$ and $0^{\circ}<l\leq90^{\circ}$, $90^{\circ}<l\leq180^{\circ}$, $180^{\circ}<l\leq270^{\circ}$, $270^{\circ}<l\leq360^{\circ}$. The halo stars with $6<R\leq10$ kpc could be observed only in the vertical distance interval $5<z\leq8$ kpc, while those with $10<R\leq15$ kpc and $15 <R\leq 20$ kpc could be separated into two sub-samples, i.e. $5<z\leq 10$, $10<z\leq 12$ kpc, and $7<z\leq10$, $10<z\leq15$ kpc respectively. The results are tabulated in Table 2. One can see that vertical metallicity gradients could not be estimated for the halo stars in the second and third quadrants for three Galactic latitudes, corresponding to the interval $6<R\leq10$ kpc, due to the insufficient number of stars. A similar case holds for the halo stars with Galactic latitude $50^{\circ}<b\leq65^{\circ}$ and, galactocentric distance $10<R\leq 15$ kpc, again, in the second and third quadrants. Also, vertical metallicity gradients could not be estimated for the thick-disc stars ($2<z\leq5$ kpc) with $80^{\circ}<b\leq90^{\circ}$ and $10<R\leq15$ kpc in the first and fourth quadrants, due to the reason just cited. The final sub-samples for which vertical metallicity gradients could not be estimated consist of halo stars with Galactic latitudes $65^{\circ}<b\leq80^{\circ}$ and $80^{\circ}<b\leq90^{\circ}$, and galactocentric distances $15<R\leq20$ kpc in the first and fourth quadrants. 

The vertical metallicity gradient could be estimated for the thin disc stars only in the galactocentric distance interval $6<R\leq 10$ kpc. There is no  systematic difference between the gradients estimated for the thin disc for different sub-samples. Correspondingly the metallicity gradients estimated for the thin disc are neither Galactic latitude nor Galactic longitude dependent. The vertical metallicity gradient $d[{\rm Fe/H}]/dz=-0.128\pm0.072$ dex kpc$^{-1}$ for the thick disc in the galactocentric distance interval $6<R\leq10$ kpc is a bit different than the counterparts. The uncertainty of this metallicity gradient is also (absolutely) larger than the counterparts. However, the cited difference can be explained by fewer number of stars used for estimation of the metallicity gradient in question. The different vertical metallicity gradients for the thick disc stars with $10<R\leq15$ kpc, such as the one of $d[{\rm Fe/H}]/dz=-0.018\pm0.080$ dex kpc$^{-1}$, are due to the fewer number of stars considered in the gradient determination. One can see a small difference between the vertical metallicity gradients, $d[{\rm Fe/H}]/dz=-0.156\pm0.038$ and $d[{\rm Fe/H}]/dz=-0.131\pm0.038$ dex kpc$^{-1}$, in the second and third quadrants, estimated for the thick-disc stars with $10<R\leq15$ kpc and $65^{\circ}< b\leq 80^{\circ}$. The first one is compatible with the counterparts in the interval $6<R\leq10$ kpc. Hence, it seems that the difference in question originates from the positions of the stars in the third quadrant. The vertical metallicity gradients estimated for the halo stars with different galactocentric distance -and Galactic coordinate- intervals are compatible, with some exceptions however where only a small number of stars are used in the estimate.

In the second procedure, the Galactic latitudes and longitudes are omitted in the definition of the sub-samples. The results are given in the third column of Table 3 and in Fig. 7. The vertical metallicity gradient for the thin disc, $0.5<z\leq2$ kpc, is high in the interval $6<R\leq10$ kpc, namely $d[{\rm Fe/H}]/dz=-0.308\pm0.018$ dex kpc$^{-1}$, while it could not be estimated at higher galactocentric distances due to the insufficient number of stars. The vertical metallicity gradients for the thick disc, $2<z\leq 5$ kpc, estimated for the intervals $6<R\leq10$ and $10<R\leq15$ kpc are compatible, $d[{\rm Fe/H}]/dz=-0.164\pm0.014$ dex kpc$^{-1}$ and $d[{\rm Fe/H}]/dz=-0.172\pm0.016$ dex kpc$^{-1}$ respectively, while it could not be carried out at distances larger than $R=15$ kpc. There is a small vertical metallicity gradient for halo stars, $5<z\leq 8$ kpc, i.e. $d$[Fe/H]$/dz=-0.023\pm0.006$ dex kpc$^{-1}$ only in the galactocentric distance interval $6<R\leq10$ kpc, while it is close to zero at larger radial distances. 

\begin{figure}[!b]
\begin{center}
\includegraphics[scale=0.52, angle=0]{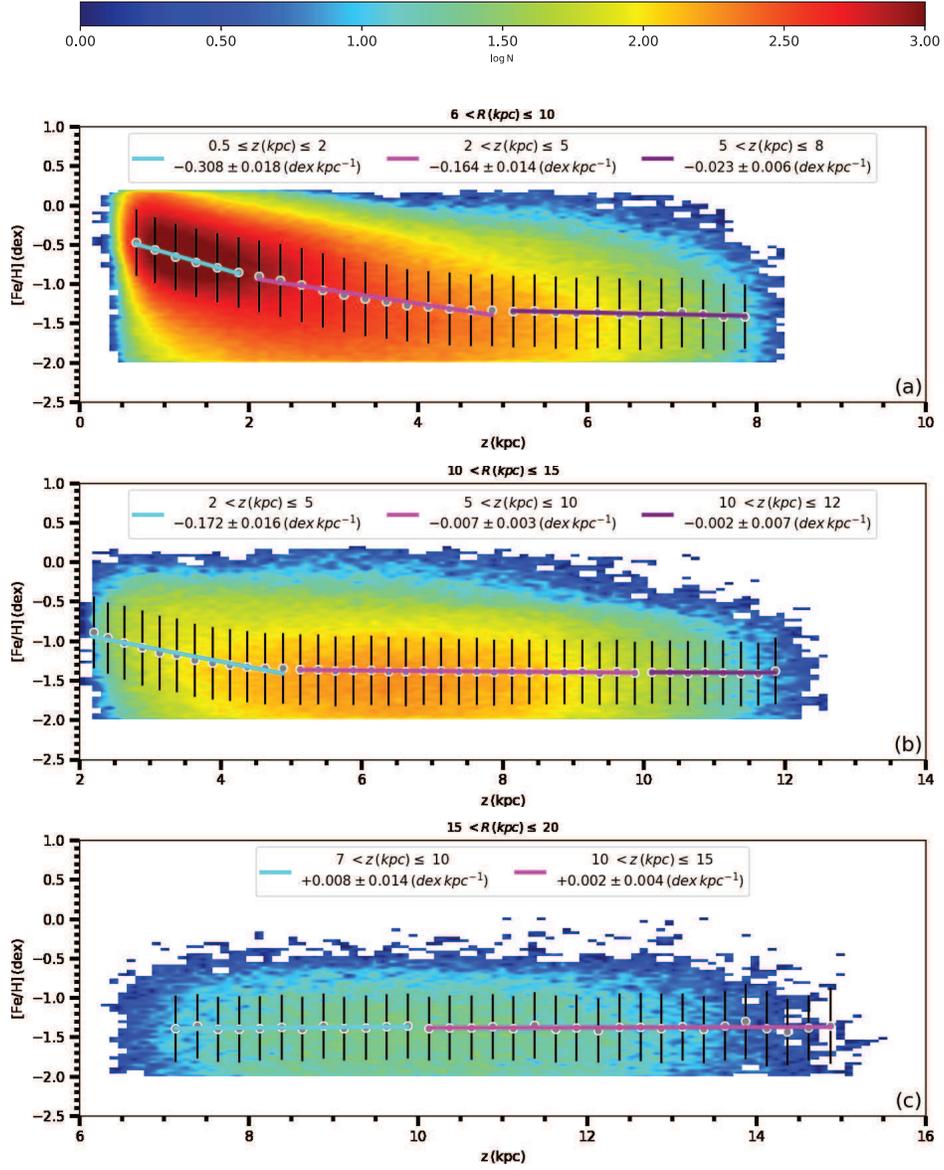}
\caption[] {Vertical metallicity gradients for three galactocentric distance intervals, (a) $6<R\leq10$, (b) $10<R\leq15$, and (c) $15<R\leq20$ kpc separated into different sub-intervals defined by vertical $z$ distances.}
\end{center}
\end{figure}

As expected, the vertical metallicity gradient in our study is highest for short vertical distances (for thin-disc stars), while, gradually it becomes lower at higher distances and diminishes at distances larger than $z=5$ kpc where halo stars dominate. This result can be confirmed by all studies in the literature. The metallicity gradient $d[{\rm Fe/H}]/dz=-0.308\pm0.018$ dex kpc$^{-1}$ estimated for the interval $0.5<z\leq2$ kpc in our study is similar to the value, $d[{\rm Fe/H}]/dz=-0.320\pm0.010$ dex kpc$^{-1}$, obtained in \citet{Yaz10} estimated for G type main-sequence stars with $z<2.5$ kpc, and likewise  our gradient $d[{\rm Fe/H}]/dz=-0.164\pm0.014$ dex kpc$^{-1}$ corresponding to the interval $2<z\leq 5$ kpc confirms the metallicity gradient in \citet{Kordopatis11}, $d[{\rm Fe/H}]/dz=-0.140\pm0.050$ dex kpc$^{-1}$, estimated for the FGK main-sequence stars with $1 \leq z\leq 4$ kpc (see Section 4 for further discussion). 

We estimated the uncertainties of the metallicity gradients by a procedure the following. We adopted the uncertainties of the metallicity and absolute magnitude cited in Paper II, i.e. $\Delta {\rm [Fe/H]}=\pm 0.137$ dex and $\Delta M_g = \pm 0.18$ mag, and produced 1,000 catalogues which cover $1.44 \times 10^6$ stars with their metallicities plus uncertainties with range $-0.137 \leq \Delta{\rm [Fe/H]} \leq +0.137$ dex, and distances plus the uncertainties estimated via their absolute magnitude uncertainties, $-0.18 \leq \Delta M_g \leq +0.18$ mag. Then, we estimated the metallicity gradients and the corresponding uncertainties for each catalogue. It turned out that the maximum uncertainty for the metallicity gradient is $\varepsilon_{\rm [Fe/H]} = \pm 0.03$ dex kpc$^{-1}$. Actually, the uncertainties for the metallicity gradients in Table 2 and Fig. 4 are less than this value. The uncertainties of the distance are estimated by means of the uncertainties of the  apparent magnitude $g$, $-0.025\leq g\leq 0.025$, and the uncertainties of the absolute magnitude $M_g$, $-0.18\leq M_g\leq +0.18$. The relative distance uncertainty is as follows: $\%8.6\leq \Delta d/d\leq \%10.2$.    

The uncertainties claimed in Section 3 are the combination of probable and systematic errors estimated as explained in the following. We fitted the [Fe/H] metallicities to the corresponding $z$ distances for each sub-sample of stars defined in the tables: ${\rm[Fe/H]}=a_0z+b_0$. The uncertainty of the inclination for  such a linear fitting ($\epsilon_1$) is the probable error for the metallicities in that sub-sample. The uncertainty for the metallicity of a single star in our study is estimated as $\epsilon_2=\pm0.03$ dex, while for the distance ($\epsilon_3$) it lies in the interval $\%8.6\leq \Delta d/d\leq \%10.2$. We reduced the distances $d$ to the $z$ ones, combined them with the corresponding uncertainties and fitted them to the metallicities which are also the combination of the original values and their uncertainties: ${\rm [Fe/H]}\pm \epsilon_2=a_1(Z\pm\epsilon_3)+b_1$. Then, we adopted the difference between two inclinations, $\epsilon_4=|a_0-a_1|$, as the systematic error of the metallicity gradients for the sub-sample in question. Then, the final uncertainty is ($\sqrt{\epsilon_1^2+\epsilon_4^2}$).  

\begin{table*}[!htb]
\scriptsize
\setlength{\tabcolsep}{2pt}
\centering
\caption{Vertical metallicity gradients for three galatocentric distance intervals, (a) $6 < R \leq 10$, (b) $10 < R \leq 15$, and (c) $15 < R \leq 20$, and 96 sub-samples, as defined in the text.} 
\begin{tabular}{ccccc|ccc|ccc}
\hline
\multicolumn{11}{c}{$6 < R \leq 10$ kpc} \\
\hline
\multicolumn{2}{c}{} & \multicolumn{3}{c}{$50^{\rm o}<b\leq 65^{\rm o}$} & \multicolumn{3}{c}{$65^{\rm o}<b\leq 80^{\rm o}$} & \multicolumn{3}{c}{$80^{\rm o}<b\leq 90^{\rm o}$}\\
\hline
$z$ & $l$ & $N$ & [Fe/H] & $d[{\rm Fe/H}]/dz$ & $N$ & [Fe/H] & $d[{\rm Fe/H}]/dz$ & $N$ & [Fe/H] & $d[{\rm Fe/H}]/dz$ \\
(kpc) & ($^{\rm o}$) &  & (dex) & (dex kpc$^{-1}$) &  & (dex) & (dex kpc$^{-1}$) &  & (dex) & (dex kpc$^{-1}$) \\
\hline
$0.5< z\leq  2$ & $  0 < l \leq  90$ & 104235 & -0.70 & -0.296$\pm$0.015 & 33928 & -0.73 & -0.312$\pm$0.023 & 5095 & -0.73 & -0.309$\pm$0.032 \\
                & $ 90 < l \leq 180$ &  63546 & -0.65 & -0.298$\pm$0.015 & 23604 & -0.70 & -0.292$\pm$0.020 & 4765 & -0.68 & -0.329$\pm$0.039 \\
                & $180 < l \leq 270$ &  64145 & -0.70 & -0.317$\pm$0.016 & 24679 & -0.72 & -0.328$\pm$0.019 & 4532 & -0.75 & -0.293$\pm$0.040 \\
                & $270 < l \leq 360$ &  71916 & -0.68 & -0.317$\pm$0.025 & 33661 & -0.69 & -0.311$\pm$0.022 & 4720 & -0.76 & -0.367$\pm$0.055 \\
$2< z \leq 5$   & $  0 < l \leq  90$ & 122960 & -1.12 & -0.166$\pm$0.016 & 44165 & -1.13 & -0.162$\pm$0.011 & 7014 & -1.13 & -0.174$\pm$0.016 \\
                & $ 90 < l \leq 180$ &  26706 & -0.98 & -0.128$\pm$0.072 & 20450 & -1.05 & -0.151$\pm$0.033 & 5621 & -1.07 & -0.171$\pm$0.017 \\
                & $180 < l \leq 270$ &  31401 & -1.03 & -0.150$\pm$0.035 & 23535 & -1.06 & -0.169$\pm$0.011 & 5705 & -1.15 & -0.180$\pm$0.015 \\
                & $270 < l \leq 360$ &  90278 & -1.11 & -0.160$\pm$0.014 & 49471 & -1.09 & -0.171$\pm$0.011 & 7127 & -1.15 & -0.158$\pm$0.021 \\
$5< z \leq 8$   & $  0 < l \leq  90$ &  34457 & -1.37 & -0.022$\pm$0.007 & 10265 & -1.37 & -0.016$\pm$0.022 &  746 & -1.37 & -0.032$\pm$0.080 \\
                & $ 90 < l \leq 180$ &    --- &   --- &              --- &   --- &   --- &              --- &    5 &   --- &              --- \\
                & $180 < l \leq 270$ &    --- &   --- &              --- &   --- &   --- &              --- &    1 &   --- &              --- \\
                & $270 < l \leq 360$ &  28982 & -1.34 & -0.027$\pm$0.006 & 12471 & -1.35 & -0.030$\pm$0.014 &  825 & -1.39 & -0.044$\pm$0.085 \\
\hline
\multicolumn{11}{c}{$10 < R \leq 15$ kpc} \\
\hline
\multicolumn{2}{c}{}& \multicolumn{3}{c}{$50^{\rm o}<b\leq 65^{\rm o}$} & \multicolumn{3}{c}{$65^{\rm o}<b\leq 80^{\rm o}$} & \multicolumn{3}{c}{$80^{\rm o}<b\leq 90^{\rm o}$}\\
\hline
$z$ & $l$ & $N$ & [Fe/H] & $d[{\rm Fe/H}]/dz$ & $N$ & [Fe/H] & $d[{\rm Fe/H}]/dz$ & $N$ & [Fe/H] & $d[{\rm Fe/H}]/dz$ \\
(kpc) & ($^{\rm o}$) &  & (dex) & (dex kpc$^{-1}$) &  & (dex) & (dex kpc$^{-1}$) &  & (dex) & (dex kpc$^{-1}$) \\
\hline

$2< z \leq 5$   & $  0 < l \leq  90$ &    542 & -1.30 & -0.018$\pm$0.080 &    19 &   --- &              --- & ---  & ---   &              --- \\
                & $ 90 < l \leq 180$ &  30248 & -1.21 & -0.171$\pm$0.014 &  6937 & -1.31 & -0.156$\pm$0.038 &  461 & -1.33 & -0.136$\pm$0.058 \\
                & $180 < l \leq 270$ &  33360 & -1.24 & -0.167$\pm$0.017 &  8065 & -1.32 & -0.131$\pm$0.038 &  464 & -1.37 & -0.196$\pm$0.077 \\
                & $270 < l \leq 360$ &    374 & -1.33 & +0.021$\pm$0.084 &    29 &   --- &              --- &  --- &   --- &              --- \\
$5< z \leq 10$  & $  0 < l \leq  90$ &  24897 & -1.38 & -0.014$\pm$0.005 & 18487 & -1.40 & -0.010$\pm$0.004 & 4283 & -1.41 & -0.016$\pm$0.008 \\
                & $ 90 < l \leq 180$ &  22253 & -1.36 & -0.028$\pm$0.025 & 17445 & -1.38 & -0.011$\pm$0.006 & 4643 & -1.39 & -0.002$\pm$0.004 \\
                & $180 < l \leq 270$ &  29162 & -1.37 & +0.010$\pm$0.013 & 20756 & -1.38 & -0.007$\pm$0.005 & 4480 & -1.41 & +0.003$\pm$0.012 \\
                & $270 < l \leq 360$ &  21172 & -1.38 & -0.000$\pm$0.005 & 22013 & -1.37 & -0.005$\pm$0.006 & 4416 & -1.40 & -0.003$\pm$0.007 \\
$10< z \leq 12$ & $  0 < l \leq  90$ &   2888 & -1.40 & -0.034$\pm$0.041 &  2814 & -1.41 & +0.016$\pm$0.042 &  487 & -1.38 & -0.039$\pm$0.076 \\
                & $ 90 < l \leq 180$ &    --- &   --- &              --- &    77 & -1.38 & -0.004$\pm$0.150 &  149 & -1.36 & -0.005$\pm$0.031 \\
                & $180 < l \leq 270$ &    --- &   --- &              --- &    82 & -1.37 & -0.003$\pm$0.032 &  133 & -1.45 & -0.002$\pm$0.081 \\
                & $270 < l \leq 360$ &   2936 & -1.38 & +0.013$\pm$0.028 &  3339 & -1.38 & -0.020$\pm$0.013 &  517 & -1.43 & -0.047$\pm$0.052 \\
\hline
\multicolumn{11}{c}{$15 < R \leq 20$ kpc} \\
\hline
\multicolumn{2}{c}{}& \multicolumn{3}{c}{$50^{\rm o}<b\leq 65^{\rm o}$} & \multicolumn{3}{c}{$65^{\rm o}<b\leq 80^{\rm o}$}& \multicolumn{3}{c}{$80^{\rm o}<b\leq 90^{\rm o}$}\\
\hline
$z$ & $l$ & $N$ & [Fe/H] & $d[{\rm Fe/H}]/dz$ & $N$ & [Fe/H] & $d[{\rm Fe/H}]/dz$ & $N$ & [Fe/H] & $d[{\rm Fe/H}]/dz$ \\
(kpc) & ($^{\rm o}$) &  & (dex) & (dex kpc$^{-1}$) &  & (dex) & (dex kpc$^{-1}$) &  & (dex) & (dex kpc$^{-1}$) \\
\hline
$7< z \leq 10$ &  $  0 < l \leq  90$ &    110 & -1.30 & -0.003$\pm$0.096 &   --- &   --- &              --- &  --- &   --- &              --- \\
               &  $ 90 < l \leq 180$ &   6863 & -1.37 & +0.012$\pm$0.024 &  1422 & -1.39 & +0.060$\pm$0.068 &   11 &   --- &              --- \\
               &  $180 < l \leq 270$ &   8944 & -1.38 & +0.010$\pm$0.016 &  1821 & -1.39 & +0.053$\pm$0.040 &    7 &   --- &              --- \\
               &  $270 < l \leq 360$ &     80 & -1.26 & +0.002$\pm$0.086 &   --- &   --- &              --- &  --- &   --- &              --- \\
$10< z \leq 15$&    $0 < l \leq  90$ &   1925 & -1.31 & +0.052$\pm$0.071 &  2101 & -1.40 & +0.014$\pm$0.013 &  536 & -1.35 & -0.007$\pm$0.028 \\
               &   $90 < l \leq 180$ &   1977 & -1.35 & -0.013$\pm$0.013 &  2957 & -1.39 & +0.020$\pm$0.013 &  898 & -1.36 & +0.001$\pm$0.021 \\
               &  $180 < l \leq 270$ &   3082 & -1.37 & -0.015$\pm$0.015 &  3983 & -1.40 & +0.011$\pm$0.015 &  850 & -1.41 & +0.023$\pm$0.022 \\
               &  $270 < l \leq 360$ &   1946 & -1.36 & +0.020$\pm$0.029 &  2577 & -1.37 & -0.000$\pm$0.009 &  602 & -1.42 & -0.023$\pm$0.026 \\
\hline
\end{tabular}
\end{table*}

\begin{table}[!htb]
\setlength{\tabcolsep}{5pt}
\centering
\caption{Vertical metallicity gradients estimated by using the observed (third column) and synthetic (fourth column, {\em Galaxia}) data.}
\begin{tabular}{cccc} 
\hline
                &                    & Observed Data      &  Synthetic Data \\
\hline
 $R$ interval   & $z$ interval       & $d[{\rm Fe/H}]/dz$ & $d[{\rm Fe/H}]/dz$\\
(kpc)           &  (kpc)             & (dex kpc$^{-1}$)   &  (dex kpc$^{-1}$) \\
\hline
$6 < R\leq 10$  & $0.5<z\leq 2$ &	$-$0.308$\pm$0.018 & $-$0.333$\pm$0.036\\
                & $2<z\leq 5$   &	$-$0.164$\pm$0.014 & $-$0.108$\pm$0.006\\
                & $5<z\leq 8$   &	$-$0.023$\pm$0.006 & $-$0.143$\pm$0.014\\
$10 < R\leq 15$ & $2<z\leq 5$   &	$-$0.172$\pm$0.016 & $-$0.233$\pm$0.015\\
                & $5<z\leq 10$  &	$-$0.007$\pm$0.003 & $-$0.034$\pm$0.005\\
                & $10<z\leq 12$ &	$-$0.002$\pm$0.007 & $+$0.006$\pm$0.008\\
$15 < R\leq 20$ & $7<z\leq 10$  &	$+$0.008$\pm$0.014 & $-$0.011$\pm$0.007\\
                & $10<z\leq 15$ &	$+$0.002$\pm$0.004 & $-$0.012$\pm$0.003\\
\hline
\end{tabular}
\end{table}

\section{Summary and Discussion}

We applied the metallicity and absolute magnitude calibrations in terms of the UV excess, $\delta_{0.41}$, presented in Paper II, and estimated vertical and radial metallicity gradients for 1,437,467 F-G type main-sequence stars with metallicities $-2\leq[{\rm Fe/H}]\leq0.2$ dex, provided from the recent survey DR12 of SDSS III \citep{Alam15}. The radial metallicity gradients are estimated for 14 distance intervals $0.44<z\leq0.87$, $0.87<z\leq 1.31$, ..., $12.21<z\leq 13.95$ kpc. The evaluation of the gradients is carried out by fitting the mean metallicities of the stars in 18 Galactic longitude intervals, $0^{\rm \circ}<l\leq 20^{\rm \circ}$, $20^{\rm \circ}<l\leq 40^{\rm \circ}$, ..., $340^{\rm \circ}<l\leq 360^{\rm \circ}$ to the corresponding median radial distances, $\widetilde{R}$. The inclination of each fit is adopted as the metallicity gradient for the star in question.   
  
The radial metallicity gradients for two intervals with short distances, $0.44<z\leq 0.87$, $0.87<z\leq 1.31$ kpc, are less than expected: $d[{\rm Fe/H}]/dR=-0.011\pm0.022$, $d[{\rm Fe/H}]/dR=-0.029\pm0.013$ dex kpc$^{-1}$, respectively. This is due to the short radial range of the corresponding stars: $7.62<R<8.42$ and  $7.50<R<8.65$ kpc. The radial metallicity gradient $d[{\rm Fe/H}]/dR=-0.041\pm0.011$ dex kpc$^{-1}$, in the interval $1.31<z\leq1.74$ kpc is close to the ones in \citet{Boeche13}: $d[{\rm Fe/H}]/dR=-0.059\pm0.002$ dex kpc$^{-1}$, and \citet{Recio14}: $d[{\rm Fe/H}]/dR=-0.058\pm0.008$ dex kpc$^{-1}$, where the first gradient is estimated for the F-G main-sequence stars with $4.5<R<9.5$ kpc, while the second one is carried out for the FGK main-sequence thin-disc stars. It is interesting that the radial metallicity gradient in \citet{Onal16}, estimated for RC stars at shorter distances $|z|<0.5$ kpc, $d[{\rm Fe/H}]/dR=-0.047\pm0.003$ dex kpc$^{-1}$ is almost the same as our gradient $d[{\rm Fe/H}]/dR=-0.041\pm0.011$ dex kpc$^{-1}$. The radial metallicity gradients for distances larger than $z=1.74$ kpc are positive or negative numbers close zero, which confirm previous findings in the literature. The study of \citet{Onal16} is an example: $d[{\rm Fe/H}]/dR=-0.001\pm0.003$ and $d[{\rm Fe/H}]/dR=+0.015\pm0.008$ dex kpc$^{-1}$ were estimated for the intervals $0.5<|z|\leq 1$ and $1<|z|\leq 3$ kpc, respectively.   

We adopted the galactocentric distance ranges $6 < R \leq 10$, $10 < R \leq 15$ and $15 < R \leq 20$ kpc and estimated vertical metallicity gradients for stars with $0.5 < z \leq 2$ (thin disc), $2 < z \leq 5$ (thick disc), and $z > 5$ (halo) kpc by using two procedures. In the first procedure the stars are separated into a series of sub-samples by using three Galactic latitude and four Galactic longitude intervals: $50^{\circ}<b\leq65^{\circ}$, $65^{\circ}<b\leq80^{\circ}$, $80^{\circ}<b\leq 90^{\circ}$ and $0^{\circ}<l\leq90^{\circ}$, $90^{\circ}<l\leq180^{\circ}$, $180^{\circ}<l\leq270^{\circ}$, $270^{\circ}<l\leq360^{\circ}$. The halo stars with $6 < R \leq 10$ kpc could be observed only in the vertical distance interval $5 < z \leq 8$, kpc, while those with $10 < R \leq 15$ kpc and $15 < R \leq 20$ kpc could be separated into two sub-samples, i.e. $5 < z \leq 10$, $10 < z \leq 12$ kpc, and $7 < z \leq 10$, $10 < z \leq 15$ kpc respectively. Vertical metallicity gradients could not be estimated for the thick disc stars with $10 < R \leq 15$  kpc and $80^{\circ}<b\leq90^{\circ}$ in the first and fourth quadrants due to a insufficient number of stars in these sub-samples. A similar case holds for the following halo stars: a) stars with $6 < R \leq 10$ kpc and $50^{\circ}<b\leq90^{\circ}$ in the second and third quadrants, b) stars with $10 < R \leq 15$ kpc, $50^{\circ}<b\leq65^{\circ}$ and $10 < z \leq 12$ kpc in the second and third quadrants, and c) stars with $15 < R \leq 20$ kpc, $65^{\circ}<b\leq90^{\circ}$ and $7 < z \leq 10$ kpc in the first and fourth quadrants.     
 
The vertical metallicity gradient could be estimated for the thin-disc stars only in the galactocentric distance interval $6 < R \leq 10$ kpc. There is no any systematic difference between the gradients estimated for the thin disc for different sub-samples. The thick disc is dominant in two galactocentric distance intervals, $6<R\leq 10$ and $10<R\leq 15$ kpc, and the gradients estimated for different sub-samples are compatible with two exceptions. The first one is related with less number of stars, such as $d[{\rm Fe/H}]/dz=-0.018{\pm 0.080}$ dex kpc$^{-1}$ in the sub-sample defined by $10 < R \leq 15$ kpc, $50^{\circ}<b\leq65^{\circ}$ and $0^{\circ}<l\leq90^{\circ}$. While the second one, i.e. $d[{\rm Fe/H}]/dz=-0.131\pm 0.038$ dex kpc$^{-1}$ estimated for stars with $10<R \leq 15$ kpc, $65^{\circ}<b\leq80^{\circ}$ and $180^{\circ}<l\leq270^{\circ}$ could not be explained. The vertical metallicity gradients estimated for the halo stars with different galactocentric distance -and Galactic coordinate- intervals are compatible, with a few exceptions where a small number of stars are used in the study.

The second set of vertical metallicity gradients is free of Galactic latitude and longitude. The high gradients cover the short vertical -and radial- distances, while they decrease gradually with increasing vertical distances and become almost zero at large distances. The highest vertical metallicity gradient $d[{\rm Fe/H}]/dz=-0.308\pm0.018$ dex kpc$^{-1}$ estimated for the thin-disc stars in our study is close to the value $d[{\rm Fe/H}]/dz=-0.305\pm0.011$ dex kpc$^{-1}$ given in \citet{Hayden14} as well as $d[{\rm Fe/H}]/dz=-0.290\pm0.060$ dex kpc$^{-1}$ in \citet{Marsakov06} estimated for red-giant stars with $0 < |z| \leq 2$ kpc and thin disc F-G type stars, respectively. The two vertical metallicity gradients estimated for the thick-disc stars for the intervals $6 < R \leq 10$ and $10 < R \leq 15$ kpc in our study, $d[{\rm Fe/H}]/dz=-0.164\pm 0.014$ dex kpc$^{-1}$ and $d[{\rm Fe/H}]/dz=-0.172\pm0.016$ dex kpc$^{-1}$ are compatible with the vertical metallicity gradients in \citet{Ak07b} and \citet{Kordopatis11}, i.e., $d[{\rm Fe/H}]/dz=-0.160\pm0.020$ dex kpc$^{-1}$ and $d[{\rm Fe/H}]/dz=-0.140\pm0.050$ dex kpc$^{-1}$ which were estimated for G type main-sequence and FGK main-sequence stars, respectively. However, the distance ranges covered by the sample stars in the cited studies are different, $z<3$ kpc (north) and $1 \leq z \leq 4$ kpc, respectively. Vertical metallicity gradients for stars at distances larger than $z=5$ kpc, where the halo component dominates, is almost zero. A case which is valid in the studies appeared in the literature.   

In summary, the vertical metallicity gradients are high at short vertical distances, while they become lower at higher vertical distances. Also, the vertical metallicity gradients of stars of different populations (dwarfs, giants, etc.) may be compatible, though their distance ranges may be different. Some vertical metallicity gradients that appeared in the literature and are cited in our study, were attributed to the three main populations of our Galaxy, i.e. thin and thick discs and halo. The highest vertical metallicities belong to the thin-disc stars very close to the Galactic plane, and the smallest ones (absolutely small numbers) were obtained at large vertical distances, i.e. $z>5$ kpc, and were attributed to halo stars. While the vertical metallicity gradients determined for the thick-disc stars, that lie between those of the thin-disc and of the halo. However, the vertical metallicity gradients and the distance ranges of the corresponding stars cited by different researchers for a given population may be different. In our study, the metallicity gradient $d[{\rm Fe/H}]/dz=-0.308\pm0.018$ dex kpc$^{-1}$ , estimated for the stars with $0.5<z\leq2$ kpc can be attributed to the thin-disc stars. While the two values, $d[{\rm Fe/H}]/dz=-0.164\pm0.014$ and $d[{\rm Fe/H}]/dz=-0.172\pm0.016$ dex kpc$^{-1}$ estimated for the stars with $2<z\leq5$ kpc correspond to the thick-disc stars. In contrast, the metallicity gradients $- 0.023\leq d[{\rm Fe/H}]/dz\leq0.008$ dex kpc$^{-1}$ estimated for stars with $z>5$ kpc belong to the halo stars.       

The three radial metallicity gradients estimated for stars with $0.44<z\leq 0.87$, $0.87<z\leq 1.31$ and $1.31<z\leq 1.74$ kpc can be attributed to thin-disc stars. The lack of a radial metallicity gradient for stars with $z>1.74$ kpc indicates that the thick disc has not undergone a radial collapse phase as observed in recent spectroscopic surveys \citep{Cheng12, Anders14, Hayden14, Recio14}.     

Our final comparison is carried out between the vertical metallicity gradients estimated in our study and those produced by {\em Galaxia} \citep{Sharma11}. {\em Galaxia} is a C++ code to generate a synthetic structure of the Milky Way for different sky surveys. Given one or more colour-magnitude diagrams, a survey size and geometry, the code returns a catalogue of stars in accordance with a given model of our Galaxy. In our case, the synthetic stars are generated for a field with size 5,280 deg$^2$, centered at the north Galactic pole. We applied the constraints presented in Section 2.1 and obtained 2,230,167 synthetic stars for our purpose. As we could not detect a strong indication for the dependence of our vertical metallicity gradients on the Galactic latitude and longitude, we restricted our comparison on the sub-samples defined by the galactocentric distance ranges $6 < R \leq 10$, $10 < R \leq 15$ and $15 < R \leq 20$ kpc. As in the observational data, the gradients are carried out for the thin disc ($0.5<z\leq 2$ kpc), thick disc ($2<z\leq 5$ kpc), and halo ($z>5$ kpc). We produced synthetic [Fe/H] metallicities and vertical $z$-distances for 2,230,167 F-G type main-sequence stars with Galactic coordinates $50^{\circ}< b \leq 90^{\circ}$ and $0^{\circ} < l \leq 360^{\circ}$ for our purpose. The results are given in the last column of Table 3 and in Fig. 8. The distribution of the (synthetic) stars corresponding to the lines in Table 3 from top to bottom is as follows: 788,679; 852,941; 142,718; 113,438; 238,559; 23,898; 24,940; and 44,994.

\begin{figure}[!b]
\begin{center}
\includegraphics[scale=0.5, angle=0]{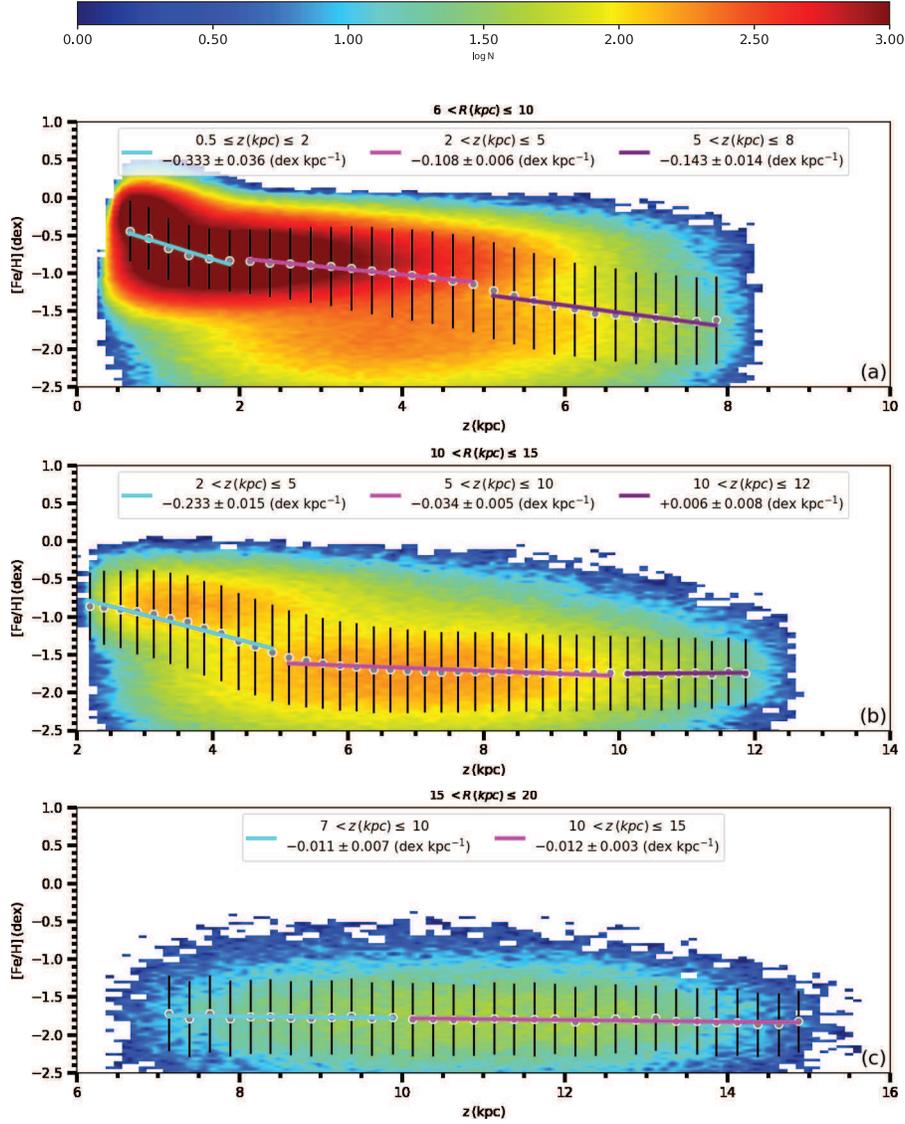}
\caption[] {Vertical metallicity gradients for three galactocentric distance intervals in the {\em Galaxia} sample, (a) $6<R\leq10$, (b) $10<R\leq15$, and (c) $15<R\leq20$ kpc. Each interval is separated into sub-intervals of $z$-distance.}
\end{center}
\end{figure}

There is a good agreement between the observed and synthetic vertical metallicity gradients for the thin disc component of our Galaxy (first line in Table 3). However, one can not confirm the same agreement for the thick disc, neither with stars with $6 < R \leq 10$ kpc (second line) nor with the ones with $10 < R \leq 15$  kpc (fourth line). It is interesting that the mentioned two observed gradients, i.e. $d[{\rm Fe/H}]/dz=-0.164\pm0.014$ dex kpc$^{-1}$ and $d[{\rm Fe/H}]/dz=-0.172\pm 0.016$ dex kpc$^{-1}$, are compatible while the synthetic counterparts are different, i.e. $d[{\rm Fe/H}]/dz=-0.108 \pm 0.006$ dex kpc$^{-1}$ and $d[{\rm Fe/H}]/dz=-0.233\pm0.015$ dex kpc$^{-1}$. Hence, we can argue that the (observed) vertical metallicity gradients estimated for the thick disc are more reliable than the synthetic ones. For the halo, we could detect only a difference between the observed and synthetic vertical metallicity gradients, i.e. $d[{\rm Fe/H}]/dz=-0.023 \pm 0.006$ dex kpc$^{-1}$, $d[{\rm Fe/H}]/dz=-0.143\pm0.014$ dex kpc$^{-1}$ (third line) which corresponds to the stars with $6 < R \leq 10$ kpc. While they are compatible in four vertical distance intervals (fifth, sixth, seventh and eight lines) at larger galactocentric distances. Finally, we should note that the observed vertical metallicity gradients estimated in our study are in agreement with those in the literature. Hence, some differences between the observed and synthetic vertical metallicity gradients just cited may originate from the ``input data'' of {\em Galaxia}.

We investigated the probable impact of the streams and substructures on our results. The spatial distribution of our sample stars in Galactic coordinates is $50^{\circ}< b \leq 90^{\circ}$ and $0^{\circ} < l \leq 360^{\circ}$. Also, they are apparently brighter than $g=20$ mag and at distances $d \leq 16$ kpc relative to the Sun. Then, there are three candidate streams, Sagittarius, Hercules-Aquila and Virgo Overdensity, which would impact our results. The spatial distributions in Galactic coordinates, their lengths and the distances of these streams are tabulated in Table 4. The distance of the Sagittarius stream cited in \citet{Ibata01}, $d=46$ kpc, is out of the distance range in our study. Hence, one can not expect any impact on our results. The impact of the same stream in \citet{Belokurov06} is limited with distance $d=15$ kpc and Galactic latitude $b=50^{\rm \circ}$. However, the Galactic longitude interval for this stream is $180^{\circ} < l \leq 230^{\circ}$, and one can not detect any  mean metallicity different than its counterparts in any third quadrant in Table 2. Hence, we can say that there is no any impact of the stream just claimed on our results. The spatial distributions of the streams Sagittarius, Hercules-Aquila, and Virgo Overdensity in \citet{Koposov12}, \citet{Simion14}, and \citet{Carlin12}, respectively, and the stars in our study have limited overlap in spatial distribution in Galactic coordinates and distance. Therefore the metallicity gradients are not contaminated by those of the stars in the cited streams. Actually, the mean metallicity, $\langle{\rm [Fe/H]}\rangle$, for stars in our study with Galactic longitude corresponding to the stream in question is compatible with those estimated for stars in different quadrants for a given vertical distance interval. This is a strong indication for our argument. As an example, we compare the mean metallicities for stars with $10 < R \leq 15$ kpc, $2 < z \leq 5$ kpc and $50^{\circ}< b \leq 65^{\circ}$  in Table 2. The mean metallicities ${\rm [Fe/H]}=-1.24$ dex and ${\rm [Fe/H]}=-1.33$ dex in the Galactic longitude intervals $180^{\circ} < l \leq 270^{\circ}$  and $270^{\circ} < l \leq 360^{\circ}$ which cover the Galactic longitude distribution $180^{\circ} < l \leq 350^{\circ}$ of the Sagittarius stream in \citet{Koposov12}, are compatible with those in the Galactic longitude intervals $0^{\circ} < l \leq 90^{\circ}$ and $90^{\circ} < l \leq 180^{\circ}$, i.e. ${\rm [Fe/H]}=-1.30$ dex and ${\rm [Fe/H]}=-1.24$ dex.

We could detect 4,819 main sequence stars with $S/N\geq50$ and $15<g<17$ mag for which SDSS spectroscopic metallicities are available in the SEGUE survey, and compared these metallicities with those determined photometrically in our study. The result is given in Fig. 9. The mean and the standard deviation of the differences between the two sets of metallicities are $\langle \Delta\rm{[Fe/H]}\rangle=0.13$ and $\sigma=0.33$ dex respectively. The large dispersion is due to the medium-resolution of the spectra \citep{Lee08a, Lee08b, Allende-Prieto08}.

We used the oblate parameter $\eta=0.85$ in our calculations. However, we tested three more values, $\eta=0.9$, 0.95 and 1, just to see the difference. The ranges of the metallicity gradients for the sub-intervals $5<Z\leq8$, $5<Z\leq10$, $10<Z\leq12$, $7<Z\leq 10$ and $10<Z\leq15$ kpc for different three radial distance intervals are $d{\rm[Fe/H]}/dz:$ [-0.023, -0.017], [-0.007, -0.006], [-0.002, 0.000], [+0.009, +0.005] and [+0.002, +0.013] dex kpc$^{-1}$, respectively. Then, we can say that a different oblate parameter would not change our results.

Improving the new metallicity and absolute magnitude calibrations by using the ultraviolet-excesses of stars  provided from the Gaia era, and their applications to the photometric data observed in the deep sky survey programs will be an important contributor for understanding of the structure, formation and evolution of the Galaxy, as well as for testing the Galactic models.
\begin{table}[!htb]
\setlength{\tabcolsep}{1.5pt}
\centering
\caption{Stellar streams in the star field used in this study. The columns give: the name of the stream, the Galactic coordinates, the extension of the stream, distance relative to the Sun, and reference.}
\begin{tabular}{cccccc}
\hline
Designation& $l$         & $b$          & $L$   & $d$   & Reference\\
       &($^{\circ}$) & ($^{\circ}$) & (kpc) & (kpc) &        \\
\hline
Sagittarius       & 350         & 50        & 50    & 46    & \citet{Ibata01}     \\
Sagittarius       & $180<l<230$ & 50        & 45    & 15    & \citet{Belokurov06} \\
Sagittarius       & $180<l<350$ & $20<b<70$ & ---   & 10-60 & \citet{Koposov12}   \\
Hercules-Aquila   & $|l| <30$   & 50        & 1-6   & 10-20 & \citet{Simion14}    \\
Virgo Overdensity & $265<l<283$ & $50<b<68$ & ---   & 14    & \citet{Carlin12}    \\
\hline
\end{tabular}%
\end{table}

\begin{figure}[!b]
\begin{center}
\includegraphics[scale=0.50, angle=0]{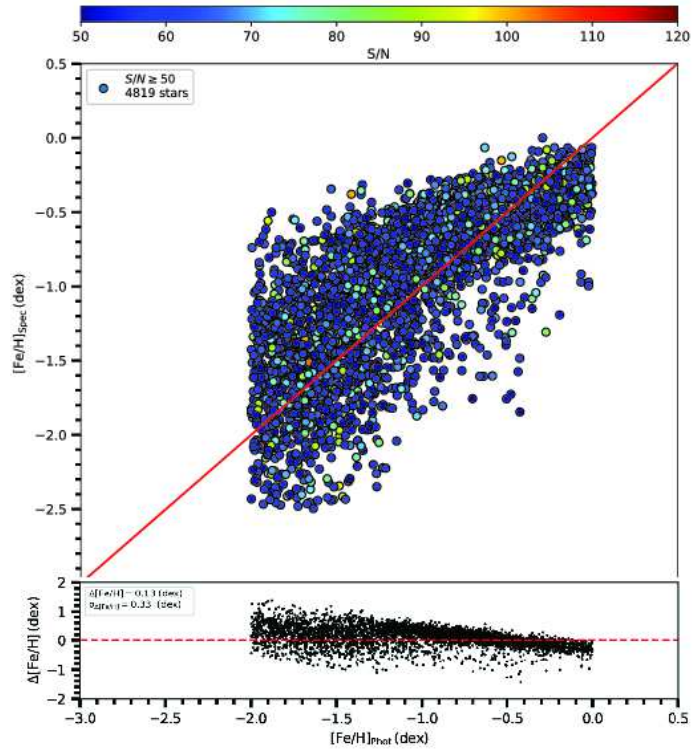}
\caption[] {Comparison of the metal abundances for 4,189 stars estimated by the photometric and spectroscopic procedures.}
\end{center}
\end{figure}

\section{Acknowledgments}
We thank the anonymous referees for their insightful and constructive 
suggestions, which significantly improved the paper.
This work has been supported in part by the Scientific and
Technological Research Council (T\"UB\.ITAK) 114F347. Part of this
work was supported by the Research Fund of the University of Istanbul,
Project Number: 52265. This research has made use of the SIMBAD, 
and NASA\rq s Astrophysics Data System Bibliographic Services. 
The authors are very grateful to Dr. Kai O. Schwenzer for his careful 
reading of the English manuscript.

\end{document}